\documentclass[twocolumn,preprintnumbers,amsmath,amssymb]{revtex4}
\usepackage{graphicx}
\usepackage{dcolumn}
\usepackage{bm}
\usepackage{comment}
\usepackage{rotating}
\usepackage{longtable}
\usepackage{float}
\usepackage{eucal}
\usepackage{csquotes} 
\usepackage{multirow}
\usepackage{booktabs}



\begin{document}

\title{\textbf{Evidence of the Antimagnetic rotation in an odd-odd nucleus: The case of $^{142}$Eu}}

\author{Sajad Ali}
\author{S. Rajbanshi}
\altaffiliation{Presently at Dum Dum Motijheel College, Kolkata 700074, India}
\author{B. Das}
\author{S. Chattopadhyay}
\author{M. Saha Sarkar}
\author{A. Goswami}
\email{asimananda.goswami@saha.ac.in}
\affiliation{Saha Institute of Nuclear Physics, HBNI, Kolkata 700064, India}
\author{R. Raut}
\affiliation{UGC-DAE-Consortium for Scientific Research, Kolkata 700098, India} 
\author{Abhijit Bisoi}
\affiliation{Indian Institute of Engineering Science and Technology, Howrah 711103, India}
\author{Somnath Nag}
\affiliation{National Institute of Technology, Raipur 492010, India}
\author{S. Saha}
\altaffiliation{Presently at GSI Helmholtzzentrum für Schwerionenforschung, Darmstadt, Hesse, Germany}
\author{J. Sethi}
\altaffiliation{Presently at Department of Chemistry and Biochemistry, University of Maryland, United States}
\author{R. Palit}
\affiliation{Tata Institute of Fundamental Research, Mumbai 400005, India}
\author{G. Gangopadhyay}
\affiliation{Department of Physics, University of Calcutta, Kolkata 700009, India}
\author{T. Bhattacharjee}
\author{S. Bhattacharyya}
\author{G. Mukherjee}
\affiliation{Variable Energy Cyclotron Center, Kolkata 700064, India}
\author{A. K. Singh}
\affiliation{Indian Institute of Technology, Kharagpur 721302, India}
\author{T. Trivedi}
\affiliation{Guru Ghasidas Vishayavidyalaya, Bilaspur 495009, India}

\date{\today}

\begin{abstract}
The present work reported a conclusive evidence for anti-magnetic rotational (AMR) band in an odd-odd nucleus $^{142}$Eu. Parity of the states of a quadrupole  sequence in $^{142}$Eu was firmly identified from polarization measurements using the Indian National Gamma Array and lifetimes of some of the states in the same structure were measured using the Doppler shift attenuation method. The decreasing trends of the deduced quadrupole transition strength $B(E2)$ with spin, along with increasing ${J}^{(2)}/B(E2)$ values conclusively established the origin of these states as arising from Antimagnetic rotation. The results were well reproduced by numerical calculations within the framework of a semi-classical geometric model.
\end{abstract}

\maketitle

It was well established that the deviation of nuclear shape from spherical symmetry results in a regular band like structures connected by quadrupole ($E2$) transitions \cite{bohr}. The quantal rotation about an axis perpendicular to the symmetric axis was possible for these nuclei. These bands were characterized by nearly constant electric quadrupole transition rates [$B(E2)$] throughout the band \cite{nolan, jankhoo, hanwu}.

However, the observation of the band-like structures in the weakly deformed Cd, Pd nuclei in mass A $\sim$ 110 region leads to the extension of the familiar concept of rotation. These bands were characterized by the decreasing $B(E2)$ rates with increasing spin while the dynamic moment of inertia [$J^{(2)}$] remains constant throughout the band \cite{hubel, roy1, roy2, simons, simons2}. These experimental signatures were well explained by the anisotropy in current distribution due to the presence of the few high spin particles and holes outside the nearly spherical core within the framework of shears mechanism \cite{hubel, mac, clark, mac-clar}. In this formalism, a special arrangement of valence particles and holes forms a double shears structure that had $\pi$ rotational symmetry about the axis of total angular momentum. At the band head these valence particles (holes) were initially aligned in time reversed orbits \cite{roy1, roy2, simons, simons2, raj3}. The high spin states along the band was generated by the simultaneous closing of the two blades of the conjugate shears, produced by the valence particles (holes). The transition rates for such a structure was expressed as $$B(E2;I \rightarrow I-2) = {\frac{15}{32\pi} {(eQ_{eff})^2}{{sin}^4{\theta}}}$$ where $Q_{eff}$ is the effective quadrupole moment of the core \cite{simons2} and $\theta$ is the angle between the particle and hole angular momentum vector with the total angular momentum axis. As the shears close $B(E2)$ transition rates will show a decreasing trend with increasing spin and consequently the ratio of dynamic moment of inertia (${J}^{(2)}$) with $B(E2)$ value rapidly increases. Since such a structure has no net transverse component of magnetic dipole moment, the M1 transition rate vanishes for these bands. Analogous to an anti-ferromagnetic substance where one half of the atomic dipole moments are aligned on one sub-lattice and the other half are aligned in the opposite direction on the other sub-lattice in a crystal structure, this phenomenon was termed as antimagnetic rotation and the resulting bands were called antimagnetic rotational (AMR) bands.

Till day, the AMR bands were observed in weakly deformed even-even and odd mass nuclei \cite{roy1, roy2, raj3, simons, simons2}, but not in an odd-odd nucleus. It was worth noting that if the anisotropy of current distribution was the primary criterion for the existence the AMR bands in the weakly deformed nuclei, there was no reason of observing them in nearly spherical odd-odd nuclei with appropriate orbital occupancy to form the double shears structure. Recently, we had reported the observation of AMR bands in the $^{143}$Eu [Z = 63, N = 79] nucleus \cite{raj3}. This was the maiden confirmation of such structure outside the A $\sim$ 100 region. In the present paper we reported an extension of this quest (for AMR) in the neighboring odd-odd $^{142}$Eu isotope. 

The $^{142}$Eu (Z = 63, N = 79) nucleus had one proton hole and three neutron holes with respect to the semi-magic nucleus $^{146}$Gd. The three neutron holes were restricted to the $h_{11/2}$ orbital. In the proton sector, the protons can be easily excited to the $h_{11/2}$ orbital across the Z = 64 sub-shell closure, leading to breaking of the anisotropy in current distribution. The nucleus $^{142}$Eu was previously studied by the Piiparinen $et$ $al$. \cite{piipar} using the $^{110}$Pd($^{37}$Cl, 4n) reaction with the NORDBALL Compton-suppressed multi-detector array facility. They reported a quadrupole band-like structure above the 1397-keV (11$^{+}$) state. In the present investigation, the aforementioned quadrupole band-structure was extended to higher spin and the parity of the states were reassigned from a complete spectroscopic measurement. In addition, the lifetimes of the states in the band were determined using the Doppler Shift Attenuation Method (DSAM). The band was interpreted as an AMR band in the light of the B(E2) values, extracted from the level lifetime and the theoretical calculations using the semi-classical particles-plus-rotor model (SPRM). This was the maiden observation of the AMR phenomenon in an odd-odd nucleus. It was further observed that the core contribution to the total angular momentum in this band gradually increases with increasing spin, an intriguing feature hitherto unobserved in similar structures established elsewhere \cite{roy1, roy2, raj3, simons, simons2}.

\begin{figure}[b]
\centering
\setlength{\unitlength}{0.05\textwidth}
\begin{picture}(10,9.5)
\put(-1.9,-0.5){\includegraphics[width=0.5\textwidth,height=0.55\textwidth]{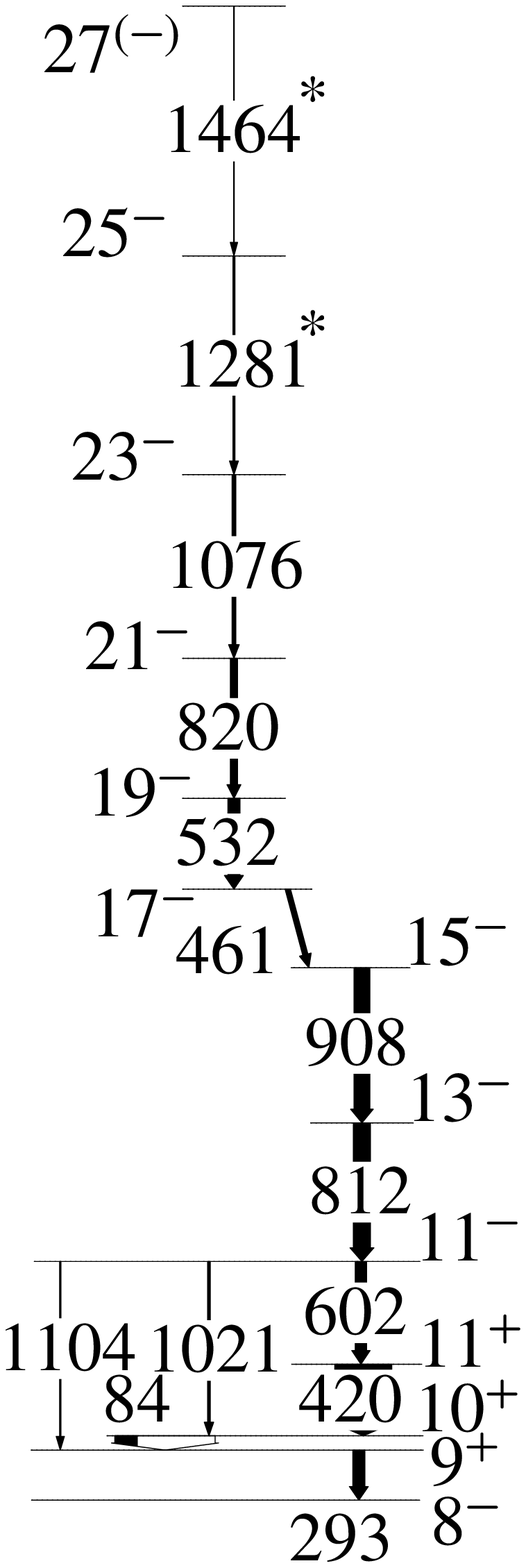}}
\put(3.7,4.4){\includegraphics[height=3.8cm]{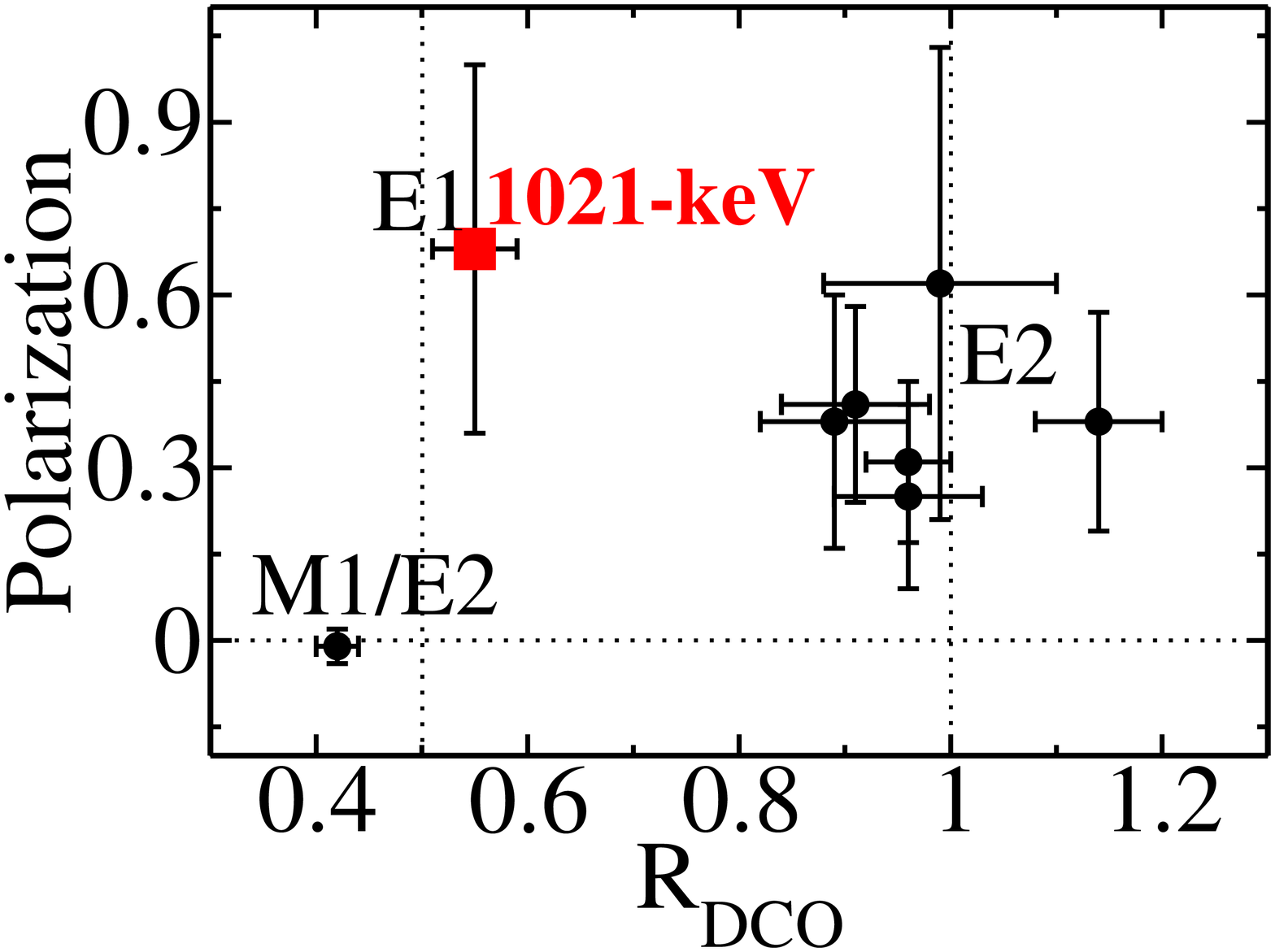}}
\put(0.8,0.3){\textbf{0}}
\put(2.0,2.9){\textbf{(a)}}
\put(7.85,7.6){\textbf{(b)}}

\end{picture}
\caption{\footnotesize (Color  online) (a) The partial level scheme of $^{142}$Eu obtained in the present work. The $\gamma$ energy are given in keV and the width of the arrows are proportional to the intensities of the transitions. The energies are rounded of to the nearest keV. Asterisk mark transition are newly observed. (e) The polarization value against DCO ratio ($R_{DCO}$) for different transitions of the observed structure in $^{142}$Eu, as shown in plot (a).}
\label{levelsc}
\end{figure}

The excited states of the nucleus of interest $^{142}$Eu were populated through the fusion evaporation reaction $^{31}$P($^{116}$Cd, 5n) at $E_{lab}$ = 148-MeV. The $^{31}$P beam obtained from the 14UD Pelletron Linac facility at TIFR, Mumbai. The target was 2.4 mg/cm$^{2}$ of $^{116}$Cd, enriched to 99$\%$, on a 14.5 mg/cm$^{2}$ thick Pb backing. The residues, produced at $\beta$ $\sim$ 2\% of c, were stopped in the target and backing. The de-exciting $\gamma$ rays were detected with the Indian National Gamma Array (INGA) then consisting of nineteen Compton-suppressed Clover detectors. The detectors were positioned at six different angles, four at 90$^{\circ}$ and three each at 40$^{\circ}$, 65$^{\circ}$, 115$^{\circ}$, 140$^{\circ}$, and 157$^{\circ}$ with respect to the beam direction. A digitizer based pulse processing and data acquisition system was used to acquire time stamped list mode data. A total of 4 $\times$ 10$^9$ two and higher fold (coincident) events were acquired during the experiment.

The time-stamped data was sorted offline using Multi-parameter time-stamped based Coincidence Search (MARCOS) program developed at the TIFR, Mumbai  \cite{palit, htan} into several symmetric and asymmetric $E_\gamma$ - $E_\gamma$  matrices as well as $E_\gamma$ - $E_\gamma$ - $E_\gamma$ cube, which were analyzed using the RADWARE and the INGASORT codes \cite{radford1, radford2, ingasort}. 

The partial level scheme of $^{142}$Eu was proposed on the basis of coincidence relationships and intensity considerations. The multipolarities and the electromagnetic character of the observed $\gamma$-ray transitions for assigning the spin parity of the levels were determined from the measurements of their angular distribution, the ratio for directional correlation from oriented state (DCO ratio)\cite{rajban}, angular distribution from oriented nuclei (ADO ratio) \cite{piipar, rajban}, and linear polarization \cite{lee, rajban}. Details of the data-analysis procedures were described in Refs. \cite{rajban, rajban2, sajad}.  

\begin{figure}[b]
\centering
\setlength{\unitlength}{0.05\textwidth}
\begin{picture}(10,8.4)
\put(1.4,4.2){\includegraphics[height=4.5cm]{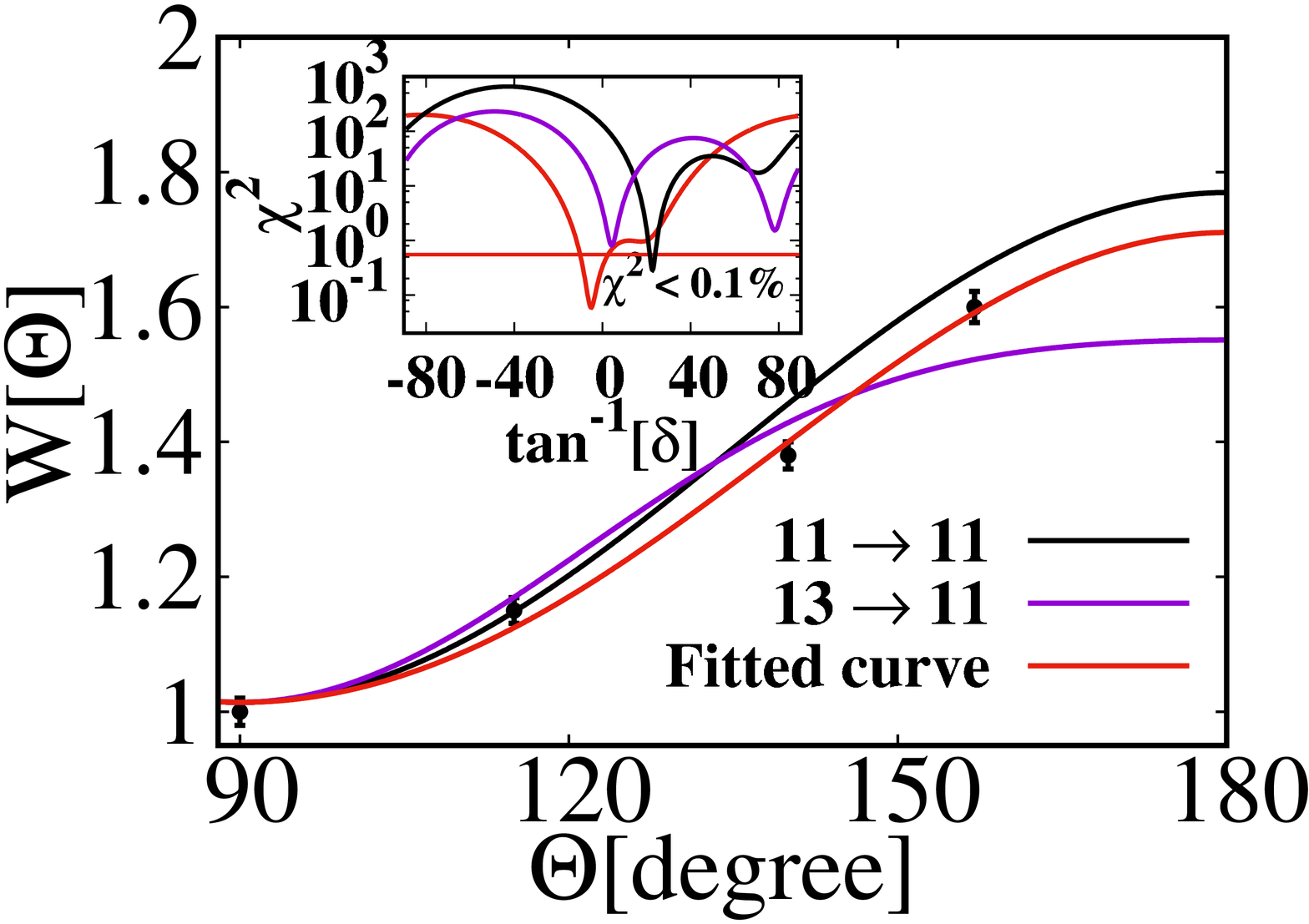}}
\put(1.4,-.3){\includegraphics[height=4.5cm]{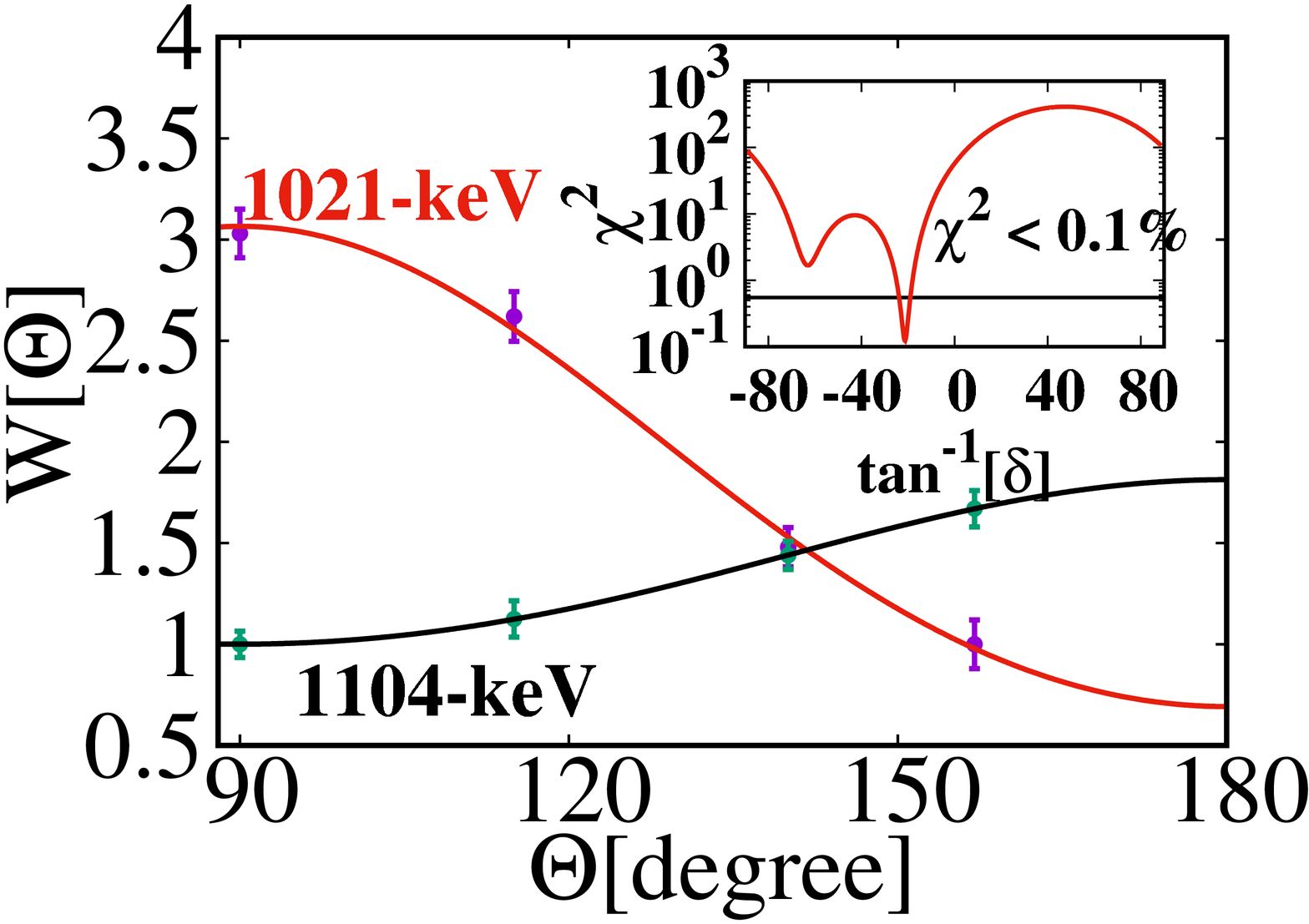}}
\put(3.0,5.8){\textbf{(a)}}
\put(3.0,3.6){\textbf{(b)}}
\end{picture}
\caption{\footnotesize (Color  online) (a) The measured angular distribution along with the fitted curve (red line) for the 601.5-keV transition. The theoretical distribution of this transition for the spin sequences 13 $\rightarrow$ 11 (violet line) and 11 $\rightarrow$ 11 (black line) are also shown. The chi-square of angular distribution coefficients for different mixing ratio ($\delta$) plotted in the inset of this figure for the spin sequence 13 $\rightarrow$ 11 (violet line), 12 $\rightarrow$ 11 (black line) and 11 $\rightarrow$ 11 (red line), respectively. (b) The experimental angular distributions  along with the fitted curves for the 1021.1 and 1104.5-keV transitions. The chi-square analysis for the experimental distribution of 1021.1-keV transition assuming the spin sequence 12 $\rightarrow$ 11 is shown in the inset of this figure and gives the mixing ratio $\delta$ = -0.39(0.01).}
\label{angdist}
\end{figure}

In the previous work by Piiparinen \textit{et al.} \cite{piipar} the 1397-keV state was found to be de-excited via the decay out 601.5, 1021.1 and 1104.5-keV $\gamma$-ray transitions. The spin-parity of the 1397-keV state was assigned 11$^+$ on the basis of evaluated ADO ratios of the depopulating transitions. In the present work, this assignment was re-examined and modified to 11$^{-}$ [Fig. \ref{levelsc} (a)]. The justification for the same was elaborated hereafter.

We had performed the angular distribution of all the transitions de-exciting the 1397-keV state by measuring their normalized yield at different angles [$W(\theta$)] and fitting with the Legendre polynomial \cite{rajban, sajad} as,  

\begin{center}
$W(\theta)= A_0[1 + {a_{2}P_{2}(cos\theta)} + {a_{4}P_{4}(cos\theta)}]$.
\end{center}

\begin{figure}
	\centering
\includegraphics[width=0.35\textwidth]{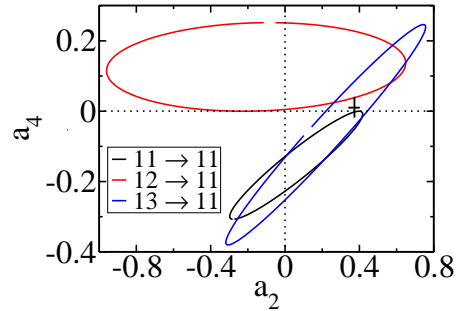}\\
\vskip -0.5cm
\caption{\footnotesize (Color  online) Angular distribution coefficients $a_2$ vs. $a_4$ plot for diffrent mixing ratio ($\delta$) using multiple spin sequence \cite{yamazaki, sajad}. The black cross indicate the data point for the 601.5-keV transition.}
	\label{a2-a4}
\end{figure}

The measured angular distribution along with the fitted curve for the three 601.5, 1021.1 and 1104.5-keV decay out transitions from the 1397-keV level were shown in Fig. \ref{angdist} (a) and (b). The extracted angular distribution coefficients $a_2$ and $a_4$ were used to determine the mixing ratio ($\delta$) of the transitions following the description as given in Ref. \cite{sajad, yamazaki}. The angular distribution of 1021.1-keV transition [Fig. \ref{angdist} (b)] exhibits a pure dipole nature with coefficients $a_2$ = -0.75(1) and $a_4$ = +0.07(6) whereas 1104.5-keV transition displays a quadrupole character with angular distribution coefficients $a_2$ = +0.42(1), $a_4$ = +0.03(1). Further, the measured values of the DCO ratio and ADO ratio for the 1021.1-keV transition were 0.55(4) and 0.53(3) respectively, indicative of a pure dipole transition contrary to the assignment in the earlier work \cite{piipar}. The polarization value for the 1021-keV transition P = 0.68(0.32) [Fig. \ref{levelsc} (b)] established it as an electric one, again at variance with the earlier result. Thus from the present measurements we propose that the 1021.1-keV transition was an $E1$ transition. This assignment requires $\Delta I$ =0, $E1$ nature for the 601.5-keV and $M2$ nature for the 1104.5-keV transitions. But due lack of statistics at 90$^{\circ}$ for the 1104.5-keV ($\Delta$ I = 2) transition, we could not determine linear polarization for the same. The characterization of the 601.5-keV transition was also crucial for upholding the 11$^{-}$ assignment of the 1397-keV state. The angular distribution coefficients of the 601.5-keV transition were $a_2$ = +0.37(3) and $a_4$ = +0.01(3) indicating it might be a $\Delta I$ = 0 or $\Delta I$ = 2 transition. To get a definitive conclusion, a contour of a$_4$ vs. a$_2$ coefficients was plotted using different spin combinations (such as 11 $\rightarrow$ 11, 12 $\rightarrow$ 11, 13 $\rightarrow$ 11) for all possible values of $\delta$ [Fig. \ref{a2-a4}]. From this contour plot it can be ascertained that the 601.5-keV transition was a $\Delta I$ = 0, dipole transition with small quadrupole admixture $\delta$ = 0.09(1). This was compliant with the $E1$ nature of the 1021.1-keV transition. The polarization measurement for the 601.5-keV transition was very difficult because of the contaminant peak structure arising from the Ge(n, n$^\prime$$\gamma$). Consequently, the 1397-keV excited state which decays through the 601.5, 1021.1 and 1104.5-keV transitions was assigned as 11$^{-}$.

Above the 1397.0-keV (11$^{-}$) state, we had observed the transitions of energies 811.7, 907.6, 460.6, 532.0, 820.1 and 1075.8-keV as proposed by the Piiparinen \textit{et al.} \cite{piipar}. The DCO ratio, ADO ratio and polarization for these were extracted in the present measurement and were in agreement with their quadrupole nature assigned previously \cite{piipar}. Two new transitions of energy 1281.0 and 1464.0-keV were also observed in the present experiment and was placed above the 23$^{-}$ state. The ADO ratio and polarization of the 1281.0-keV transition indicate its $E2$ character. For the 1464.0-keV only the ADO ratio estimation was possible and the same indicates it to be a quadrupole transition. In the absence of the polarization data, we had tentatively assign the spin parity of the state depopulating by 1464.0-keV transition as 27$^{(-)}$.

\begin{figure}[b]
\centering
\setlength{\unitlength}{0.05\textwidth}
\begin{picture}(10,7.5)
\put(-.6,0){\includegraphics[height=7.6cm]{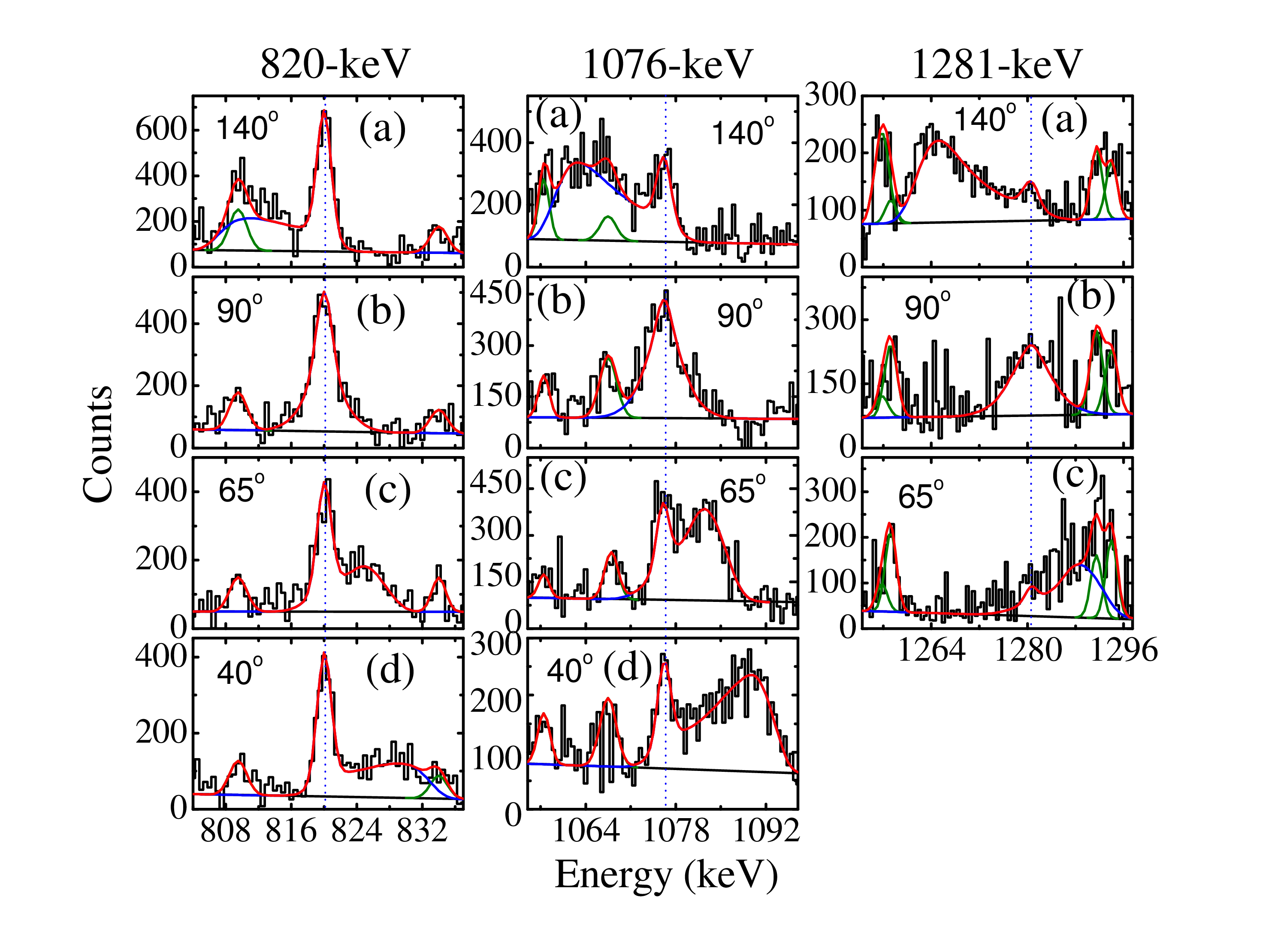}}
\end{picture}
\vskip -0.5cm
\caption{\footnotesize (Color online) The experimental observed spectra along with the fitted line shapes for the 820.1, 1075.6 and 1281.0-keV transitions of the quadrupole band of $^{142}$Eu. The figures (a), (b), (c) and (d) are correspond to the shapes in $140^{\circ}$, $90^{\circ}$, $65^{\circ}$ and $40^{\circ}$ detectors. Shape at $40^{\circ}$ spectra for 1281-keV merges with background. The obtained line shape of the $\gamma$ transition, contaminant peaks and the total line shapes are represented by the blue, olive and red curves, respectively. The vertical dotted line represents the stopped peak position for each transition, whose shape is observed.}
	\label{lineshape}
\end{figure}

\begin{figure}[t]
\centering
\setlength{\unitlength}{0.05\textwidth}
\begin{picture}(8,9.5)
\put(-1.4,9.6){\includegraphics[width=0.5\textwidth,height=0.63\textwidth,angle=-90]{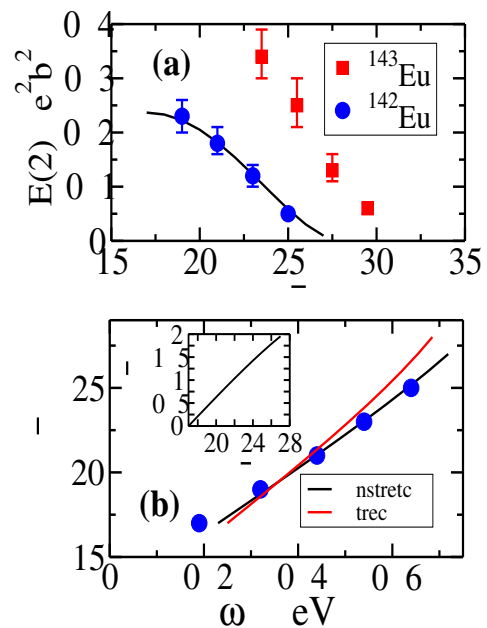}}
\end{picture}
\caption{\footnotesize (Color online) (a) $B(E2)$ values with spin [I($\hbar$)] (b) I($\hbar$) against rotational frequency ($\omega$) for the quadrupole band in $^{142}$Eu. The variation of $R$ is shown in the inset plot. The black and red line represents the theoretical SPRM calculation for unstretch and stretch condition, respectively. The parameter used for this calculation are $V_{\pi\pi}$ = 0.2 and $V_{\pi\nu}$ = 1.7-MeV \cite{raj3}, $j_\pi^1$ = $j_\pi^2$ = 4.5 [unstretch], $j_{\nu}$ = 17, n = 8 and a = 3.78 and $eQ_{eff}$ = 1.26 eb (calculated as in Ref. \cite {mac-clar}). The parameter haad the same meaning as in Refs. \cite{roy1, roy2, simons, simons2, raj3}. The upward arrow in (a) represents the lower limit of $B(E2)$ value for the 25$^-$ state of the quadrupole structure. The $B(E2)$ values for the quadrupole band I in $^{143}$Eu \cite{raj3} are also shown (red squares) (see text for details). }
	\label{BE2}
\vskip -0.6cm
\end{figure}

\begin{table*}
  
\caption{ Measured level lifetimes, the corresponding $B(E2)$ transition rates, dynamic moment of inertia ${J}^{(2)}$ and the ratio of ${J}^{(2)}/B(E2)$ values for the quadrupole transitions in \text{$^{142}$Eu}.}
\begin{ruledtabular}
\begin{tabular}{cccccccc}

\text{$J_{i}$$^{\pi}$}  & \text{$E_\gamma$} & \text{${\tau}$\footnote{The level lifetimes from Ref. \cite{piipar}}}  &  \text{${\tau}$\footnote{Present measurements}}  & \text{B(E2)} & \text{${J}^{(2)}$} & \text{${J}^{(2)}/{B(E2)}$}  \\

[\text{$\hbar$}] & [\text{keV}] &  [\text{ps}] &  [\text{ps}] &  [\text{${e^2}{b^2}$}] &  [\text{${\hbar}^{2}{MeV}^{-1}$}] &  [\text{${\hbar}^{2}{MeV}^{-1}$}\text{$/$}{\text{${e^2}{b^2}$}}] \\

\hline

\\
19$^{-}$   &  532.0  & 8.30(12)  &				 & 0.23$^{+0.03}_{-0.03}$     & 15.40 &   66.95$^{+8.73}_{-8.73}$\\
\vspace{.1cm}
21$^{-}$   &  820.1  & $<$2	   &  	1.23$^{+0.21}_{-0.16}$	 & $0.18^{+0.03}_{-0.02}$     & 16.73 &   92.94$^{+15.49}_{-10.33}$\\
\vspace{.1cm}
23$^{-}$   &  1075.8 & 		   &  	 0.46$^{+0.09}_{-0.07}$	 & $0.12^{+0.02}_{-0.02}$     & 18.98 &   158.17$^{+26.36}_{-26.36}$\\
\vspace{.1cm}
25$^{-}$   &  1281.0 & 		   & 	 $<$0.45		 &      $>$0.05		     & 20.72 & $<$414.7\\

\label{table2}
\end{tabular}
\end{ruledtabular}

\end{table*}

Doppler-broadened line shapes were observed for the three transitions depopulating $I^\pi$ = 21$^-$, 23$^-$ and 25$^-$ states of the quadrupole structure in $^{142}$Eu, following which the lifetime of these could be determined in the present work using the DSAM technique. The level lifetime of 8.3(12) ps for the 19$^-$ state was reported in the earlier work \cite{piipar} using the plunger technique, was adopted in the present study. 

The lifetime analysis in the present work was carried out using the LINESHAPE code \cite{well-john}. The stopping process of the $^{142}$Eu (residual) nuclei in the target-backing medium was simulated using Monte Carlo based approach in the code in time steps of 1.2 fs. The electronic stopping powers were calculated from the shell-corrected tabulations of Northcliffe and Schilling \cite{lcnor} and the nuclear stopping powers from the theory of Linhard, Schraff, and Schiott \cite{linha}. The LINESHAPE program uses the velocity profiles of the residues to calculate the Doppler shape for a given $\gamma$ transition. The lifetime of the corresponding state was then extracted from fitting the calculated shape to the experimental one. The process of line-shape fitting was described in detail in Refs. \cite{rajban, rajban2, sajad}.

The experimental gated spectra at 40$^{\circ}$, 65$^{\circ}$, 90$^{\circ}$ and 140$^{\circ}$ were fitted simultaneously to determining the level lifetimes. The gates were set on the transitions below the band which required incorporation of the side-feeding contribution. The same was modeled with a cascade of five transitions characterized by the same moment of inertia as that of the band under consideration \cite{nrjonson2}. Starting from the topmost transition of the band, the members of the band were sequentially fitted. After having fitted all the transitions of the band, sequentially, a global least-squares minimization was carried out for all the transitions of the cascade simultaneously, wherein only the transition quadrupole moments and the side-feeding quadrupole moments for each state were kept as free parameters. To find out the effect of side-feeding on the evaluated lifetimes, we varied the side-feeding intensities between two extreme values (by taking the corresponding uncertainties in intensities into account). The effect of variation in the side-feeding intensity resulted in a change in the level lifetime by less than 10\% \cite{rajban, sajad}. For the estimation of lifetime of 21$^{−}$ state, the contribution of the observed side-feeding transitions were also taken into account.

Lineshape fit to the observed Doppler shapes for the transitions of the quadrupole band in $^{142}$Eu were shown in Fig. \ref{lineshape}. The level lifetimes and the extracted $B(E2)$ values for the quadrupole structure were tabulated in Table \ref{table2}. The uncertainties in the lifetimes were determined from the nature of ${\chi}^2$ in the vicinity of its minimum value. The systematic errors due to the uncertainty in the stopping power of the target/backing medium, which can be as large as 15\%, were included in the quoted errors on the level lifetimes.

Variation of the quadrupole transition strengths $B(E2)$, as depicted in Fig. \ref{BE2} (b), shows a decreasing trend with increasing spin for the quadrupole structure in $^{142}$Eu. In addition, the $J^{(2)}$/$B(E2)$ values of Table \ref{table2} were found to be an order of magnitude larger than those for a well deformed collective rotor [$\sim$ 10 MeV$^{-1}$$(eb)^{-2}$] and increased with spin. This was expected for an AMR band as the $B(E2)$ values were small and decrease with spin while $J^{(2)}$ remains nearly constant \cite{hubel, roy1, roy2, raj3, simons, simons2}. The trends of the $B(E2)$ values and the $J^{(2)}$/B(E2) ratios were the definitive experimental signatures for the AMR phenomenon. Therefore, the quadrupole structure in $^{142}$Eu can be interpreted as an AMR band. This was the first conclusive observation of the AMR phenomenon in an odd-odd nucleus in the entire nuclear chart. 

The ground state (8$^{-}$) configuration for $^{142}$Eu was [$\pi({{d_{5/2}/g_{7/2}})^{-1}} \otimes \nu{h_{11/2}^{-3}}$] \cite{piipar}. In the previous work on $^{142}$Eu \cite{piipar}, the low-lying states upto 797-keV (11$^+$) were interpreted in the shell model framework with [$\pi{h_{11/2}^{+1}} \otimes \nu{h_{11/2}^{-1}}$] configuration. The 11$^-$ state at 1397-keV was expected to had a configuration as [${\pi}g_{7/2}^{-1}{\nu}h_{11/2}^{-3}$]. The anti-symmetric state with three neutron holes in $h_{11/2}$ orbital can produce a spin of $15/2$ which then coupled with a proton hole in $g_{7/2}$ orbital can produce a spin-parity of 11$^-$ with the configuration [${\pi}g_{7/2}^{-1}{\nu}h_{11/2}^{-3}$]. This particular configuration will create a favorable situation for the generation of angular momentum through shears mechanism that were observed in the several neighboring nuclei with oblate deformation \cite{rajban, rajban2, raj3, pod, past}. We propose a configuration of [${\pi}g_{7/2}^{-1}{\nu}h_{11/2}^{-3} \otimes {\pi}h_{11/2}^{2}$] for the 17$^-$ (3578-keV) state in the quadrupole band in the present work. Here the holes were rotational aligned and the particles were in the time reversed $h_{11/2}$ orbitals which were anti-aligned to each other and perpendicular to the total angular momentum of the rotation-aligned holes. Thus with this configuration, a conjugate shears structure exist and higher angular momentum states may be generated due to gradual closing of the anti-aligned angular momentum vector of particles in $h_{11/2}$ orbital, leading to an AMR band. Such alignment of the particles and holes along the deformation axis and rotation axis, respectively, requires oblate deformed nuclear shape of $^{142}$Eu. Such oblate deformed nuclear shape was observed for the shears bands in $^{142}$Sm, $^{141, 143}$Eu and in $^{142}$Gd \cite{rajban, raj3, sajad, pod, past} and interpreted in the framework of shears mechanism with the principal axis cranking and by the tilted axis cranking models.

In order to explore underlying theoretical structure of the  AMR band with the aforesaid configuration, a analysis was carried out using the Semi-classical Particle plus Rotor Model (SPRM model) \cite{mac, clark, mac-clar} based on the competition between the shears mechanism and core rotation which contain the residual interaction of particle and hole. The experimental $B(E2)$ rates and the $I(\omega$) values were reproduced simultaneously within the framework of the SPRM model using the configuration ${\pi}g_{7/2}^{-1}{\nu}h_{11/2}^{-3} \otimes {\pi}h_{11/2}^{2}$ as shown in Fig. \ref{BE2} (a). At the band-head, the two proton particles were aligned in the time-reversed $h_{11/2}$ orbital tha initially leads to j$_{\pi}$ as 5.5$\hbar$ assuming a stretched configuration of the angular momentum vectors. However, with j$_\pi$ = 5.5$\hbar$, j$_{\nu}$ = 17$\hbar$, V$_{\pi\pi}$ = 0.2 and V$_{\pi\nu}$ = 1.7 (V$_{\pi\pi}$ and V$_{\pi\nu}$ being chosen from the systematics of this mass region \cite{raj3}), it was observed that the calculation could not self-consistently reproduce both the $I(\omega)$ and $B(E2)$ values as shown in Fig. \ref{BE2} (b) (represented by solid red line). Thus the same were repeated with the proton blades in an unstretched configuration with j$_\pi$ = 4.5$\hbar$ that well reproduced the experimental results ($I(\omega)$ and $B(E2)$ values) simultaneously (Figs. \ref{BE2} (a) and (b)). Such unstretched configurations were also assumed in semi-classical calculations for other nuclei in this mass region \cite{rajban, rajban2, raj3, sajad, pod, past}. Therefore, the theoretical calculations within the the SPRM model upholds the interpretation of the band in the odd-odd $^{142}$Eu nucleus resulting from the AMR phenomenon. This was the maiden observation of the phenomenon in the odd-odd systems.

The present unstretched semi-classical calculations for the quadrupole band not only reproduced the experimental $I(\omega)$ and $B(E2)$ values simultaneously (Figs. \ref{BE2} (a) and (b)) but also established the 1464.0-keV transition as a member of the AMR band. This was understandable from the $\sim$ 2$\hbar$ core angular momentum ($R$) contribution at the top of the band (27$^{-}$) as seen from the $R$ vs. $I$ plot (inset of Fig. \ref{BE2} (b)). This core angular momentum coupled to the maximum attainable spin (25$\hbar$) for the unstretched situation of this configuration (because of Pauli-blocking) generates the observed 27$^{(-)}$ state within the AMR mechanism. Such increase of core contribution along the band results in the slow falling of the $B(E2)$ rates as in comparison with the fully stretched configuration of $^{143}$Eu as shown in Fig. \ref{BE2} (a) \cite{raj3}. Though, the competition between the collective rotation and the shears mechanism (AMR phenomenon) was observed in $^{110}$Cd nucleus \cite{roy1}, smooth increase of core contribution along the AMR band was also a solitary example, though such effect was common in the case of magnetic rotational band structure \cite{mac, clark, mac-clar, rajban, rajban2, sajad}.

In summary, the high spin quadrupole structure in $^{142}$Eu was investigated using the reaction $^{116}$Cd($^{31}$P, 4n) at a beam energy of 148 MeV. The parity of the band structure were re-assigned from the complete spectroscopy measurements. The level lifetimes in the quadrupole structure was measured. The rotational band with a band head spin of 17$^-$ was proposed as an AMR band from the falling trend of $B(E2)$ values with spin. The falling rate of the $B(E2)$ values for the AMR band in $^{142}$Eu was slow compared to the neighboring nucleus exhibiting the AMR band \cite{raj3}. The reason for the slow falling may be due to gradual increase of the core contribution along the AMR band. Numerical calculation in the framework of SPRM model well reproduce the experimental results based on the proposed configuration. The result established that the quadrupole structure above the 17$^-$ state in $^{142}$Eu was generated due to the shears mechanism manifested as AMR band. This was the first conclusive observation of the AMR phenomenon in an odd-odd nucleus. 

The authors gratefully acknowledge the financial support by the Department of Science \& Technology (DST) for the INGA project (No. IR/S2/PF-03/2003-II). We would like to acknowledge the help from all INGA collaborators. We are thankful to the TIFR-BARC Pelletron staff for giving us steady and uninterrupted $^{31}$P beam. G. G acknowledges the support provided by the University Grants Commission, Departmental Research Support (UGC-DRS) Program (No.F.530/16/DRS-II/2015(SAP-I)).

\end{document}